\documentstyle[preprint,aps,prd]{revtex}
\input{prepictex}
\input{pictex}
\input{postpictex}

\begin{document}

\draft

\preprint{SNUTP-93/80}

\title{Heavy Baryons as Skyrmion with $1/m_Q$ Corrections}

\author{Yongseok Oh$^{a)}$\footnote{Present address:
        Department of Physics, National Taiwan University,
        Taipei, Taiwan 10764, R.O.C.},
        Byung-Yoon Park$^{b)}$, and Dong-Pil Min$^{a)}$}

\address{$^{a)}$ Center for Theoretical Physics and
         Department of Physics\\
         Seoul National University, Seoul 151--742, Korea\\
         $^{b)}$ Department of Physics,
         Chungnam National University\\
         Daejeon 305--764, Korea}

\maketitle

\begin{abstract}
We take into account the $1/m_Q$ corrections upto $1/N_c$ order
in the heavy-meson-soliton
bound state approach for heavy baryons.
With these corrections, the mass spectra of baryons with
$c$-quark as well as of those with $b$-quark are well reproduced.
For charmed baryons, however, the correction to the mass spectra
amounts to about 300 MeV, which is not small compared to the
leading order binding energy, $\sim 800$ MeV.
\end{abstract}
\pacs{PACS number: 12.39.Dc, 12.39.Hg, 14.20.Lq, 14.20.Mr}

\narrowtext

\section{Introduction}
The bound-state-approach advocated by Callan and
Klebanov(CK)\cite{CK} has been shown to work very well for static
properties\cite{CKs} of strange baryons. In CK approach,
strange baryons are described by  properly quantized states of
the $K$-meson(s)-soliton bound system. Rho {\it et al\/}.\cite{RRS}
extended the CK approach further to baryons containing a heavy flavor
such as charm or bottom. In particular, the mass spectra and
magnetic moments\cite{OMRS} for charmed baryons are found to be
strikingly close to the predictions of the quark model description.
In these calculations, vector meson fields such as $K^*$, $D^*$
and $B^*$ are eliminated in favor of a combination of a background
and corresponding pseudoscalar fields, $K$, $D$ and $B$. This
approximation is valid only when vector mesons are sufficiently
heavier than corresponding pseudoscalar mesons as in the case of
$\rho$ and $\pi$ ($m_\rho^{}=770$~MeV, $m^{}_{\pi^0}=135$~MeV:
$m^{}_{\pi^0}/m_\rho^{}=0.18$). For charmed mesons or bottom
flavored mesons, however, vector mesons are only a few percent
heavier than corresponding pseudoscalar mesons: $m_{K^*}^{}=892$~MeV,
$m^{}_{K^0}=498$~MeV ($m^{}_{K^0}/m_{K^*}^{}=0.56$),
$m_{D^*}^{}=2010$~MeV, $m^{}_{D^0}=1865$~MeV
($m^{}_{D^0}/m_{D^*}^{}=0.93$), and $m_{B^*}^{}=5325$~MeV,
$m^{}_{B^0}=5279$~MeV ($m^{}_{B^0}/m_{B^*}^{}=0.99$).
Thus it is needed to treat heavy vector mesons correctly on the
same footing as heavy pseudoscalar mesons.

The heavy quark symmetry is a new spin and flavor symmetry of QCD
in the limit of infinite heavy quark masses. As a heavy quark
becomes infinitely heavy, the dynamics of a heavy quark in QCD
depends only on its velocity and is independent of its mass and spin.
This symmetry can be seen in weak semileptonic decays \cite{IW89},
mass splittings and partial decay widths \cite{IW91} of heavy mesons
and heavy baryons, whose masses are much bigger than the  QCD scale,
$\Lambda_{\mbox{\tiny QCD}}$.
Recently, effective heavy meson Lagrangians which have both
chiral symmetry and heavy quark symmetry have been constructed
by several authors\cite{BD,Wise,Yan}.
Also, a lot of works on heavy baryons as skyrmions {\it \`a la\/}
Callan-Klebanov have been reported\cite{JMW,MOPR,NRZ,GMSS}.
In a series of papers\cite{JMW}, Jenkins {\it et al.\/}
investigated the binding of a heavy meson with a soliton
using such an effective Lagrangian.
Nowak {\it et al.\/}\cite{NRZ} studied the heavy quark symmetry
in heavy baryon mass spectra in connection with the Berry's phase.
Gupta {\it et al.\/}\cite{GMSS} discussed roles of light
vector meson degrees of freedom such as $\omega$ and $\rho$.
In these works, however, only the leading order terms in the inverse
of the heavy quark mass have been considered. Also,
bound heavy mesons are assumed to sit at the center of the soliton
with their  wavefunctions taken as $\delta$-functions.
As a result, heavy mesons appear to be too deeply bound and
$\Sigma_Q$ and $\Sigma^*_Q$ are degenerate in mass.

In order to investigate more realistic cases with hyperfine
splittings, one needs to include next to leading order terms
in $1/m_Q$. In Ref.\cite{JMW}, mass corrections are roughly
estimated by including  mass differences
between heavy pseudoscalar mesons and heavy vector mesons,
while keeping the $\delta$-function-like wavefunctions.
Although it may work well for bottom flavored baryons,
we may have some doubts on the validity of such $\delta$-function-like
wavefunctions for charmed baryons: the finite mass corrections
need to be included in the wavefunctions of
heavy mesons, leading to different radial functions,
though sharply peaked at the center of the soliton.
In this paper, we attempt to establish a ``smooth" connection between
the CK approach for light baryons and the heavy-meson-soliton bound
state approach for heavy baryons by clarifying the
above-mentioned problems. In our calculation, heavy pseudoscalar
mesons and heavy vector mesons are treated on the same footing
and the next to leading order terms in $1/m_Q$, are incorporated
properly.

In the following section, we introduce a simple Lagrangian which
is relevant to our purpose. Then, a soliton-heavy-meson bound state
is found in Sec.~\mbox{III} by solving the equations of motion for the
classical eigenmodes of heavy mesons moving in the soliton
background. In Sec. IV, we discuss the mass formula for heavy baryons
containing a heavy quark.  We also discuss
the heavy quark symmetry breaking by the Wess-Zumino term
in the heavy baryon mass spectra. Section~V contains summary and
conclusion.

\section{Model Lagrangian}
In order to avoid any unnecessary complications,
we work with a simple Lagrangian for the interaction
of light Goldstone bosons with  heavy mesons, which
has the $SU(2)_L\times SU(2)_R$ chiral symmetry
and the heavy quark symmetry in the heavy mass limit.
One may obtain such a Lagrangian from the Skyrme model
Lagrangian by trimming away all the higher derivative terms
or from the heavy quark effective Lagrangian
by including the next to leading order terms in $1/m_Q$.

Up to a single derivative on the Goldstone boson fields, the most
general chirally invariant Lagrangian density\footnote{One may
improve the model Lagrangian by including terms with more
derivatives on the Goldstone boson fields and incorporating vector
mesons such as $\rho$ and $\omega$\cite{GMSS}.}
can be written
in a form of \cite{Yan}
\widetext
\begin{eqnarray}
\renewcommand\arraystretch{1.5}\begin{array}{l}
{\cal L} = {\cal L}_M + ( D_\mu \Phi)^\dagger D^\mu \Phi
- M^2_\Phi \Phi^\dagger \Phi
- \textstyle\frac12 \Phi^{*\dagger}_{\mu\nu} \Phi^{*\mu\nu}
+ M^2_{\Phi^*} \Phi_\mu^{*\dagger} \Phi^{*\mu} \\
\hskip 1cm + f_{_Q} \left( \Phi^\dagger A^\mu \Phi_\mu^*
+ \Phi_\mu^{*\dagger} A^\mu \Phi \right) + \textstyle\frac12
g_{_Q} \epsilon^{\mu\nu\lambda\rho} ( \Phi^{*\dagger}_{\mu\nu}
A_\lambda \Phi^*_\rho + \Phi^{*\dagger}_\rho
A_\lambda \Phi^*_{\mu\nu} ),
\end{array}\label{lag}
\end{eqnarray}
\narrowtext
\noindent \noindent
where $\Phi$ and $\Phi^*_\mu$ are the heavy pseudoscalar and the
heavy vector meson
doublets\footnote{Here, 
we adopt a different convention for $\Phi$ and
$\Phi^*_\mu$ than that of Ref.\cite{Yan}. Our $\Phi(\Phi^*_\mu)$
corresponds to their $\Phi^\dagger(\Phi_\mu^{*\dagger})$.
}
with masses $M_\Phi$ and $M_{\Phi^*}$, respectively.
For example, in the case of charmed mesons, we have
$$\Phi=\left(\!\!\begin{array}{c} \bar{D}^0
\\ D^- \end{array}\!\!\right),
\hskip 5mm \Phi^*=
\left(\!\!\begin{array}{c} \bar{D}^{*0} \\
D^{*-} \end{array}\!\!\right). $$
The Lagrangian density for the Goldstone boson fields is
$${\cal L}_M =
\frac{f^2_\pi}{4} Tr ( \partial_\mu U^\dagger \partial^\mu U)
+\frac{1}{32e^2} Tr [U^\dagger\partial_\mu U,
U^\dagger\partial_\nu U]^2,
\eqno(\mbox{\ref{lag}a})$$
with
$$\renewcommand\arraystretch{1}
 U\equiv \xi^2=\exp \left(\frac{i}{f_\pi}
\left(\!\begin{array}{cc} \pi^0 \!&\! \sqrt2\pi^+ \\
\sqrt2\pi^- \!&\! -\pi^0 \end{array}\!\right) \right)
\eqno(\mbox{\ref{lag}b})$$
and $f_\pi$ being the pion decay constant. The ``Skyrme term" with a
dimensionless parameter $e$ is included to stabilize the soliton
solution. Here $f_{_Q}$ and $g_{_Q}$ are the
$\Phi\Phi^*\pi$ and $\Phi^*\Phi^*\pi$ coupling constants.
The vector and axial vector potentials $V_\mu$ and $A_\mu$ are
defined in terms of $\xi$ as
$$\renewcommand\arraystretch{1.5}\begin{array}{l}
A_\mu = \frac{i}{2}(\xi^\dagger \partial_\mu \xi
- \xi \partial_\mu \xi^\dagger ),  \\
V_\mu = \frac{1}{2}(\xi^\dagger \partial_\mu \xi
+ \xi \partial_\mu \xi^\dagger ),
\end{array}\eqno(\mbox{\ref{lag}c})$$
and the covariant derivative $D_\mu$
and the field strength $\Phi^*_{\mu\nu}$ are
$$ D_\mu = \partial_\mu + V_\mu, \hskip 7mm
\Phi^*_{\mu\nu} = D_\mu \Phi_\nu^* - D_\nu \Phi_\mu^*.
\eqno(\mbox{\ref{lag}d})$$
Under $SU(2)_L\times SU(2)_R$ chiral transformations,
the fields transform as
\begin{eqnarray}
\begin{array}{c}
\xi\rightarrow \xi^\prime = L \xi h^\dagger
= h \xi R^\dagger \hskip 0.6cm
(U \rightarrow U^\prime = LUR^\dagger), \\
\Phi \rightarrow \Phi^\prime = h \Phi, \hskip 5mm
\Phi^*_\mu \rightarrow \Phi^{*\prime}_\mu = h \Phi^*_\mu,
\end{array}
\end{eqnarray}
where $L$ and $R$ are global transformations in $SU(2)_L$ and
$SU(2)_R$ respectively and $h$ is a special unitary matrix depending
on $L$, $R$ and the Goldstone fields. Furthermore, the Lagrangian
is invariant under the parity operation
\begin{eqnarray}
\begin{array}{c}
U(\vec{r},t) \rightarrow
{\cal P} U {\cal P}^{-1} = U^\dagger(-\vec{r},t), \\
\Phi(\vec{r},t) \rightarrow {\cal P} \Phi {\cal P}^{-1}
= -\Phi(-\vec{r},t), \\
\Phi^*_\mu(\vec{r},t) \rightarrow {\cal P}\Phi^*_\mu{\cal P}^{-1}
= -\Phi^*_\mu(-\vec{r},t).
\end{array}
\end{eqnarray}
Here, we have used the fact that pions and heavy mesons
(both pseudoscalar mesons and vector mesons) carry negative
intrinsic parity.

We have four  parameters in the Lagrangian to be fixed; the pion
decay constant $f_\pi$, the Skyrme parameter $e$ and
the coupling constants $f_{_Q}$ and $g_{_Q}$.
The pion decay constant, $f_\pi$, and the Skyrme parameter, $e$,
are fixed by fitting the nucleon and delta masses in the
$SU(2)$ sector\cite{ANW}.
As for the heavy meson coupling constant $f_{_Q}$ and $g_{_Q}$,
little has been known except the upper bound\cite{ACCMOR}. Thus,
we use the heavy quark symmetry as a guide line.
We use the empirical masses for the heavy meson masses, $M_\Phi$ and
$M_{\Phi^*}$. In the heavy mass limit, we have $M_\Phi \simeq
M_{\Phi^*}$ with the mass difference being of order $1/m_Q$ at most
and the two coupling constants $f_{_Q}$ and $g_{_Q}$ become related
to each other by\cite{Yan,MOPR2}
\begin{eqnarray}
f_{_Q} = 2M_{\Phi^*} g_{_Q}
\label{rel}
\end{eqnarray}
due to the heavy-quark spin symmetry. Furthermore, $g_{_Q}$
approaches a universal constant $g$ due to the heavy-quark
flavor symmetry.

The nonrelativistic quark model estimate of $g$ ($-0.75$)\cite{Yan}
is consistent with the experimental value
($|g|^2$\raisebox{-0.6ex}{$\stackrel{\textstyle <}{\sim}$} 0.5)
measured via the $D^*$ decay width\cite{ACCMOR}
and the $D^{*+}\rightarrow D^+\pi^0$
and $D^{*+}\rightarrow D^0\pi^+$ branching ratios\cite{CLEO}.

One may determine the two coupling constants $f_{_Q}$ and $g_{_Q}$
from a low energy chiral theory. In Ref.\cite{SMNR}, a Lagrangian
for the interactions of $K$ and $K^*$ mesons with pions is derived
on the basis of $SU(3)$ chiral symmetry along the hidden gauge
symmetry scheme. Comparing it with our Lagrangian, we get
$f_{_Q}/2M_{K^*}=-\frac{1}{\sqrt2}\sim -0.71$, which is very close
to the nonrelativistic quark model prediction.
Although the $\Phi^*\Phi^*\pi$ term proportional to $g_{_Q}$
is missing in Ref.\cite{SMNR}, one can find such a term among the
homogeneous solutions of the Wess-Zumino anomaly equation.
(See Ref.\cite{MOPR2} for further details.)
Using the vector meson dominance hypothesis and the empirical value
on the $g_{K^*\pi\pi}$ coupling constant ($\sim 6$), we obtain
$g_{_Q}\sim -0.7$ from the chiral Lagrangian of Ref.\cite{MOPR2}.

\section{Soliton-Heavy Meson Bound State}
The Lagrangian density ${\cal L}_M$ supports a stable $SU(2)$
soliton solution of ``hedgehog" type
\begin{equation}
U_0(\vec{r}) = \exp ( i \vec\tau \cdot \hat r F(r) ),
\label{hedgehog}
\end{equation}
with
$$F(0)=\pi \hskip 5mm\mbox{and}\hskip 5mm
 F(r)\stackrel{r\rightarrow\infty}{\longrightarrow}0.
\eqno(\mbox{\ref{hedgehog}a})$$
The above solution carries a nontrivial winding number due to its
nontrivial topological structure identified as the baryon number
$$\begin{array}{rl}
B\!\!\! &= \displaystyle \frac{1}{24\pi^2}\int d^3r\;
\varepsilon^{ijk}
 Tr(U_0^\dagger\partial_i U_0^{} U_0^\dagger\partial_j U_0^{}
     U_0^\dagger\partial_k U_0^{}) \\  \hskip 1cm
&=\displaystyle -\frac{2}{\pi} \int^\infty_0\!\!\!r^2dr
\frac{\sin^2\!F}{r^2}F^\prime =1,
\end{array}\eqno(\mbox{\ref{hedgehog}b})$$
and a finite mass
\widetext
$$ M_{sol} = 4\pi \int^\infty_0\!\!\!r^2dr \left\{
\frac{f_\pi^2}{2} ( F^{\prime 2} + 2\frac{\sin^2\!F}{r^2})
 + \frac{1}{2e^2} \frac{\sin^2\!F}{r^2}
(\frac{\sin^2\!F}{r^2} + 2F^{\prime 2}) \right\},
\eqno(\mbox{\ref{hedgehog}c})$$
with $F^\prime=\frac{dF}{dr}$.

\narrowtext

Now, our problem is to find the eigenmodes of the heavy mesons
moving in the static potentials provided by the $B=1$ soliton
configuration (\ref{hedgehog}) sitting at the origin; {\it viz.},
\begin{eqnarray} \begin{array}{l}
V^\mu=(V^0,\vec{V})=(0,i\upsilon(r)\hat{r}\!\times\!\vec\tau), \\
A^\mu=(A^0,\vec{A})=(0,\frac12(a_1(r)\vec{\tau}+a_2(r)
\hat{r}\vec\tau \cdot \hat r)),
\end{array}\end{eqnarray}
with
\begin{eqnarray}
&& \upsilon(r)=\frac{\sin^2(F/2)}{r}, \nonumber \\
&& a_1(r)=\frac{\sin F}{r} \hskip3mm\mbox{and}\hskip3mm
a_2(r)=F^\prime-\frac{\sin F}{r}.
\end{eqnarray}
The equations of motion can be read off from the
Lagrangian (\ref{lag}):
\begin{eqnarray}
( D_\mu D^\mu + M_\Phi^2 ) \Phi = f_{_Q} A^\mu \Phi^*_\mu,
\label{eq_k}
\end{eqnarray}
for the pseudoscalar meson field $\Phi$ and
\begin{eqnarray}
D_\mu \Phi^{*\mu\nu} + M^2_{\Phi^*} \Phi^{*\nu}
= - f_{_Q} A^\nu \Phi + g_{_Q} \varepsilon^{\mu\nu\lambda\rho}
A_\lambda \Phi^*_{\mu\rho}
\label{eq_phi*}
\end{eqnarray}
for the vector meson fields $\Phi^*_\mu$.

The conjugate momenta to the meson fields $\Phi$ and
$\Phi^*_\mu$ are
\begin{eqnarray}
\Pi &=& \frac{\partial {\cal L}}{\partial (\dot \Phi)} =
( D_0 \Phi)^\dagger, \nonumber \\
\Pi^{*i} &=& \frac{\partial {\cal L}}{\partial (\dot {\Phi_i^*})} =
(\Phi^{*i0})^\dagger - g_{_Q} \epsilon^{ijk} \Phi^{*\dagger}_k A_j,
\end{eqnarray}
respectively,
and we get similar equations for $\Pi^\dagger$ and $\Pi^{*i\dagger}$.
Since $\Pi^*_0$ vanishes identically, the $\Phi^*_0$ cannot be an
independent dynamical variable. We eliminate the complementary
$\Phi^*_0$ field by using Eq. (\ref{eq_phi*})
\begin{eqnarray}
\Phi^{*0} = - \frac{1}{M_{\Phi^*}^2} ( D_i \Pi^{*i\dagger}
 + \textstyle \frac12 g_{_Q} \epsilon^{ijk} A_k \Phi^*_{ij}) ,
\label{time-comp}
\end{eqnarray}
which results in a set of coupled equations
\widetext
\begin{eqnarray}\begin{array}{c}
\displaystyle
\dot{\vec \Phi}{}^* = - {\vec \Pi}^{*\dagger}
- g_{_Q} \vec A \times {\vec \Phi}^*
+ \frac{1}{M^2_{\Phi^*}}
\vec D ( \vec D \cdot {\vec \Pi}^{*\dagger} )
+ \frac{g_{_Q}}{M^2_{\Phi^*}}
\vec D( \vec A \cdot ( \vec D \times {\vec \Phi}^* )) , \\
\displaystyle
\dot {\vec \Pi}{}^{*\dagger} = \vec D \times
( \vec D \times {\vec \Phi}^* )
+ M^2 {\vec \Phi}^* + f_{_Q}
\vec A \Phi -  g_{_Q} \vec A \times {\vec \Pi}^{*\dagger}
- g_{_Q}^2 \vec A \times ( \vec A
\times {\vec \Phi}^* ) \\
\displaystyle
 - \frac{2g_{_Q}}{M_{\Phi^*}^2} \vec A \times \vec D \left\{
\vec D \cdot {\vec \Pi}^{*\dagger} +
g_{_Q} \vec A \cdot ( \vec D \times {\vec \Phi}^* ) \right\}.
\end{array}\label{Pi-dot}
\end{eqnarray}
where $\vec D = \vec \nabla - \vec V$.

\narrowtext

In order to express the equations
of motion only in terms of $\Phi$ and $\vec{\Phi}^*$,
we use the fact that the
$\Phi^*_0$ field is of order  $1/m_Q$ at most; {\it viz.\/},
\begin{equation}
\Phi^{*0} \sim \frac{1}{M^2_{\Phi^*}} D_i \dot{\Phi}^{*i}
\sim O(1/M_{\Phi^*}).
\end{equation}
Keeping this leading order term leads us to the equations of motion
\begin{eqnarray}
\ddot{\vec\Phi}{}^* &=& - 2 g_{_Q} \vec A \times \dot{\vec \Phi}{}^*
- \vec D\times ( \vec D \times {\vec \Phi}^*)
- M^2_{\Phi^*} {\vec \Phi}^* \nonumber \\
&& - f_{_Q} \vec A \Phi + \vec D ( \vec D \cdot {\vec \Phi}^*).
\label{14}\end{eqnarray}

Because of the spin-isospin mixing in the hedgehog configuration of
the classical background,
the equations of motion (\ref{eq_k}) and (\ref{14}) are invariant
only under the rotation by the grand spin $\vec{K}$ defined by
$\vec{K}=\vec{S}+\vec{I}+\vec{L}$ with $\vec{S}(\vec{I})$
being the spin (isospin) of the heavy mesons and $\vec{L}$ the
orbital angular momentum. Thus, eigenmodes are classified
by the quantum numbers $k$, $m_k$ and $P$ (the parity,
$P=(-1)^{\ell+1}$ with $\ell$ being the orbital angular momentum) as
\begin{eqnarray}\begin{array}{l}
\displaystyle
\Phi(\vec{r},t)=\sum_{k,m_k,P} \varphi_{k,m_k,P}^{}(r,t)
{\cal Y}_{k,P,m_k}^{}(\hat{r}), \\
\displaystyle
\vec{\Phi}^*(\vec{r},t)=\sum_{k,m_k,P,\kappa}
\varphi_{k,m_k,P}^{*\kappa}(r,t)
\vec{\cal Y}_{k,P,m_k}^{(\kappa)}(\hat{r}),
\end{array}\end{eqnarray}
where ${\cal Y}_{k,P,m_k}$ and $\vec{\cal Y}_{k,P,m_k}$
are the generalized spherical spinor and vector harmonics,
respectively, and $\kappa$ is an index to label the possible vector
spherical harmonics with the same $k$, $m_k$ and $P$.
To avoid cumbersome notation, we will suppress the trivial indices
$k$, $m_k$ and $P$ of the radial functions as
$\varphi(r,t)$ and $\varphi^{*\kappa}(r,t)$.

{}From now on, we will restrict our consideration to $k^P={\frac12}^+$
states, which are expected to have at least one bound state.
Since pseudoscalar mesons do not carry spin, we have only one
spherical spinor harmonics with $k^P={\frac12}^+$:
\begin{eqnarray}
{\cal Y}_{\frac12,+,\pm\frac12}(\hat{r})
 = \frac{1}{\sqrt{4\pi}} \vec\tau \cdot \hat r \chi_\pm.
\end{eqnarray}
Here, $\chi_\pm$ is the isospin basis for the heavy meson
doublets, {\it i.e.\/},
\begin{equation}
\chi_{+} = \left(\!\!\begin{array}{c} 1 \\
  0 \end{array}\!\!\right)
  \hskip 3mm\mbox{and}\hskip 3mm
  \chi_{-} = \left(\!\!\begin{array}{c} 0 \\
  1 \end{array}\!\!\right).
\end{equation}
For vector mesons with spin 1, we can construct two different
$k^P={\frac12}^+$ vector spherical harmonics\cite{fnt}:
{\it viz.\/},
\begin{eqnarray}\begin{array}{l}
\displaystyle
\vec{\cal Y}_{\frac12,+,\pm\frac12}^{(1)}(\hat{r})
= \frac{1}{\sqrt{4\pi}} \hat{r} \chi_{\pm},  \\
\displaystyle
\vec{\cal Y}_{\frac12,+,\pm\frac12}^{(2)}(\hat{r})
= i\frac{1}{\sqrt{8\pi}} (\vec{\tau}\times\hat{r}) \chi_{\pm}.
\end{array}\end{eqnarray}
Putting
\begin{eqnarray}
\Phi(\vec{r},t)
&=& \varphi(r)e^{-i\omega t}
{\cal Y}_{\frac12,+,\pm\frac12}(\hat{r}), \\
\vec{\Phi}^*(\vec{r},t)
&=& \varphi^*_1(r) e^{-i\omega t}
\vec{\cal Y}^{(1)}_{\frac12,+,\pm\frac12}(\hat{r})  \nonumber \\
&& + \varphi^*_2(r) e^{-i\omega t}
\vec{\cal Y}^{(2)}_{\frac12,+,\pm\frac12}(\hat{r}),
\end{eqnarray}
into the equations of motion (\ref{eq_k}) and (\ref{14}), we obtain
three coupled differential equations for the radial functions:
\widetext
\begin{eqnarray} \begin{array}{l}
\displaystyle
\varphi'' + \frac{2}{r} \varphi'
+ (\omega^2 - M^2_\Phi - \frac{2}{r^2}) \varphi   =
2 \upsilon(\upsilon - \frac{2}{r} ) \varphi
 + \frac{f_{_Q}}{2} (a_1+a_2) \varphi^*_1 - \frac{1}{\sqrt2}{f_{_Q}}
a_1 \varphi^*_2 , \\
\displaystyle
\varphi^{*\prime\prime}_1 + \frac{2}{r} \varphi^{*\prime}_1
+ (\omega^2 - M^2_{\Phi^*} - \frac{2}{r^2}) \varphi^*_1
=  \frac{f_{_Q}}{2} ( a_1 + a_2 ) \varphi
+ 2  \upsilon^2 \varphi^*_1 \\
\displaystyle \hskip 65mm
 + \sqrt2 ( g_{_Q} a_1 \omega - \frac{1}{r} \upsilon +
\upsilon' ) \varphi^*_2 , \\
\displaystyle
 \varphi^{*\prime\prime}_2 + \frac{2}{r} \varphi^{*\prime}_2
 + (\omega^2 - M^2_{\Phi^*} - \frac{2}{r^2} ) \varphi^*_2
=  - \frac{f_{_Q}}{\sqrt2} a_1 \varphi + \sqrt2
( \omega g_{_Q} a_1
- \frac{1}{r} \upsilon + \upsilon') \varphi^*_1   \\
\displaystyle \hskip 65mm
+ ( - \omega g_{_Q} (a_1+a_2)
- \frac{4}{r} \upsilon + 4 \upsilon^2 ) \varphi^*_2. \\
\end{array}\end{eqnarray}
\narrowtext
The wavefunctions are normalized such that each mode carries one
corresponding heavy flavor number:
\widetext
\begin{eqnarray}
1 = \int^\infty_0\!\! r^2dr \left\{
2\omega \left[|\varphi|^2 + |\varphi^*_1|^2 + |\varphi^*_2|^2 \right]
+ g_{_Q} \left[ (a_1+a_2)|\varphi^*_2|^2 - \sqrt2
a_1 ( \varphi_1^{*\dagger} \varphi^*_2
  + \varphi_2^{*\dagger} \varphi^*_1 ) \right]
\right\},
\label{norm}\end{eqnarray}
where we have kept terms up to the next to leading order in $1/m_Q$.

\narrowtext
Near the origin, the equations of motion behave asymptotically as
\begin{eqnarray}
\renewcommand\arraystretch{1.5}\begin{array}{c}
\displaystyle
   \varphi'' + \frac{2}{r}\varphi' = 0 , \\
\displaystyle
   \varphi^{*\prime\prime}_1 + \frac{2}{r}\varphi^{*\prime}_1
     - \frac{4}{r^2} \varphi^*_1
   = -\frac{2\sqrt2}{r^2} \varphi^*_2 , \\
\displaystyle
   \varphi^{*\prime\prime}_2 + \frac{2}{r}\varphi^{*\prime}_2
- \frac{2}{r^2} \varphi^*_2
   = -\frac{2\sqrt2}{r^2} \varphi^*_1.
\end{array}
\end{eqnarray}
They imply that we have three independent solution sets as
\begin{eqnarray}
\mbox{(a)} &\hskip 0.2cm&
\varphi(r) = \varphi(0) + O(r^2) ,\nonumber \\
&&\varphi^*_i(r) = O(r^2) ,\nonumber \\
\mbox{(b)} &&
\varphi(r) = O(r^2) ,\nonumber \\
&&\varphi^*_i(r) = \varphi^*_{bi}(0) + O(r^2),\nonumber \\
\mbox{(c)} &&
\varphi(r) = O(r^4) , \nonumber \\
&&\varphi^*_i(r) =
\frac12\varphi^{*\prime\prime}_{ci}(0) r^2 + O(r^4),
\end{eqnarray}
with $\sqrt2\varphi^*_{b1}(0)=\varphi^*_{b2}(0)$
and $\varphi^{*\prime\prime}_{c1}(0)
=-\sqrt2\varphi^{*\prime\prime}_{c2}(0)$.
For sufficiently large $r (\gg 1/M_\Phi)$,
the three equations decouple from each other: for example,
\begin{equation}
\varphi'' + \frac{2}{r} \varphi'
  + (\omega^2 - M_{\Phi^*}^2) \varphi = 0.
\end{equation}
Thus the bound state solutions ($\omega < M_\Phi$) are
\begin{eqnarray}
\varphi(r) &=& \alpha\frac{e^{-r\sqrt{M_{\Phi}^2-\omega^2}}}{r},
\nonumber \\
\varphi^*_1(r) &=&
\alpha_1\frac{e^{-r\sqrt{M_{\Phi^*}^2-\omega^2}}}{r}, \nonumber \\
\varphi^*_2(r) &=&
\alpha_2\frac{e^{-r\sqrt{M_{\Phi^*}^2-\omega^2}}}{r},
\end{eqnarray}
with three constants $\alpha$, $\alpha_1$ and $\alpha_2$.

The lowest energy bound states are found numerically,
and the results are shown in Table~I and Fig.~1.
In Table~I, the input parameters are listed together with the
numerical results on the lowest bound states.
In Fig. 1,  we give the radial functions $\varphi(r)$ and
$\varphi^*_1(r)$ for the $D$ and $D^*$ mesons (solid curve) and
the $B$ and $B^*$ mesons (dashed curves). By comparing the
two cases, one can easily check that as the meson mass becomes
larger, (1) the radial function becomes more sharply peaked at
the origin and (2) the role of the vector mesons becomes important
so that the radial function $\varphi^*_1(r)$ becomes comparable
to $\varphi(r)$ (see also the ratio $\varphi^*_1(0)/\varphi(0)$).
The radial function  $\varphi^*_2(r)$, though not shown in Fig. 1,
is hardly distinguishable from $\sqrt2\varphi^*_1(r)$.
This can be understood as follows:
due to their heavy masses, heavy mesons are localized in the region
$r$\raisebox{-0.6ex}{$\stackrel{\textstyle <}{\sim}$}$1/M_\Phi$,
where
\begin{eqnarray}\begin{array}{l}
[a_1(r)+a_2(r)] \sim [- a_1(r)] \sim F'(0)+O(r^2), \\
\upsilon(r) \sim \displaystyle\frac{1}{r}
-\textstyle\frac14 F^{\prime2}(0)r+\cdots,
\end{array}\label{AppPot}\end{eqnarray}
so that the equation of motion for
$(\varphi^*_1-\frac{1}{\sqrt2}\varphi^*_2)$
is completely decoupled from those for $\varphi$ and
$(\varphi^*_1+\sqrt2\varphi^*_2)$.

It would be interesting to compare our radial functions with
those of Ref.\cite{RRS} and Ref.\cite{JMW}. In Ref.\cite{RRS},
vector mesons are assumed to be sufficiently heavy and
the following {\em ansatz} is made:
\begin{eqnarray}
\Phi_\mu^* = \frac{\sqrt2}{M^{}_{\Phi^*}} A_\mu \Phi,
\label{Ansatz}
\end{eqnarray}
which implies that
\begin{eqnarray}
\varphi^*_1(r) &=& \frac{1}{\sqrt2M^{}_{\Phi^*}}
(a_1(r) + a_2(r)) \varphi(r),
\nonumber \\
\varphi^*_2(r) &=& -\frac{1}{M^{}_{\Phi^*}} a_1(r) \varphi(r).
\label{Comp}
\end{eqnarray}

As $\sqrt2\varphi_1^* \sim \varphi_2^*$ for heavy mesons due to
Eq. (\ref{AppPot}), we have only to compare $\varphi_1^*$ with
$\varphi$ in Eq. (\ref{Comp}). In the heavy mass limit, both
should play equally important roles. But the ansatz strongly
suppresses the role of vector mesons by a factor of
$\sqrt2 ef_\pi/M_{\Phi^*}$, since one obtains
$F^\prime(0)\sim - 2e f_\pi$ in the Skyrme-term-stabilized soliton
solution.  For example, this factor amounts to
0.56, 0.25 and 0.09 for the cases of $M_{K^*}$(892 MeV),
$M_{D^*}$(2010 MeV) and $M_{B^*}$(5325 MeV), respectively.
Therefore, the ansatz of Eq. (\ref{Ansatz}) is not valid
unless the vector meson is much heavier than the corresponding
pseudoscalar meson.

The wavefunctions of Refs.\cite{JMW,MOPR2} are obtained in the
heavy mass limit, $M_\Phi^{},M_{\Phi^*}^{}\rightarrow\infty$ and
can be written in our convention as
\begin{eqnarray}\renewcommand{\arraystretch}{1.5}\begin{array}{l}
\displaystyle \Phi \sim \frac12 \frac{1}{\sqrt{2M_{\Phi^*}}}
f(r) {\cal Y}_{\frac12,+,\pm\frac12}, \\
\displaystyle \vec\Phi^* \sim -\frac12\frac{1}{\sqrt{2M_{\Phi^*}}}
f(r) (\vec{\cal Y}^{(1)}_{\frac12,+,\pm\frac12}
 + \sqrt2 \vec{\cal Y}^{(2)}_{\frac12,+,\pm\frac12}),
\end{array} \label{IMlimit}\end{eqnarray}
where  the radial function $f(r)$, normalized as
$\int\!\!r^2dr |f|^2 = 1$,  is strongly peaked at the
origin. It implies that
$$ \varphi(r) = -\varphi^*_1(r) = -\frac{1}{\sqrt2} \varphi^*_2(r)
\sim \textstyle \frac12\frac{1}{\sqrt{2M_{\Phi^*}}} f(r).
\eqno(\mbox{\ref{IMlimit}a}) $$
These radial functions satisfy the normalization condition of
Eq. (\ref{norm}) in the leading order in $1/m_Q$; {\it viz.\/},
$$  2\omega_B^{} \int^\infty_0\!\! r^2dr
(|\varphi|^2 + |\varphi^*_1|^2 + |\varphi^*_2|^2 ) = 1.
\eqno(\mbox{\ref{IMlimit}b})$$
It is interesting to note that the pseudoscalar meson and three
vector mesons contribute equally to the bound state.

Comparing our numerical results given in Table I with the binding
energy $E_b=-\frac32g_{_Q} F^\prime(0)$ of Refs.\cite{JMW,MOPR2}
which gives $\sim 800$ MeV with the same input parameters, one
can see that the $1/m_Q$ corrections amount to $\sim 200$ MeV
in the bottom sector and $\sim 300$ MeV in the charm sector.
This is one of the main results of this work.

In Ref.\cite{JMW}, the rms radii of the heavy flavor current in
heavy baryons are essentially zero. Due to the $1/m_Q$ corrections,
however, we have non-zero finite size rms radii in our calculation,
{\it viz.\/} $\sim 0.3$ fm for bottom flavored baryons and
$\sim 0.4$ fm for charmed baryons.
This implies that the rms radii of heavy flavored baryons become
small as the masses become large. Due to this
effect, the binding energy is smaller than the one obtained with
$\delta$-function-type solutions.

\section{Heavy Baryons and Hyperfine Splittings}
So far we have considered soliton-heavy-meson bound states
to the order $N_c^0$ with $N_c$ being the number of color.
The combined system of the soliton and a bound heavy meson
carries a baryon number and a heavy flavor number, but does not
have the spin and isospin of a heavy baryon.
Up to order $N_c^0$, the soliton-heavy-meson bound state
 should be understood as a mixed state
of three degenerate heavy baryons containing a heavy quark $Q$;
$\Sigma_Q$, $\Lambda_Q$ and $\Sigma^*_Q$,
whose mass is $M_{sol}+\omega_B$.
In order to give the spin and isospin quantum numbers and the
hyperfine splittings, we have to go to the next order in $1/N_c$,
{\it i.e.}, $O(N_c^{-1})$.
This is done by quantizing the zero modes associated
with the simultaneous $SU(2)$ rotation of the combined system.
A standard collective coordinate quantization procedure
leads us to the mass formula for a heavy baryon
with spin $J$ and isospin $I$:
\begin{eqnarray}
M &=& M_{sol} + \omega_B^{} \nonumber \\
&& + \frac{1}{2{\cal I}} \left( c J(J+1)
+ (1-c) I(I+1) + \frac34 c(c-1)
\right) \nonumber \\
&& + O(1/M^2).
\label{mf}
\end{eqnarray}
Here ${\cal I}$ is the moment of inertia of the soliton
configuration against the $SU(2)$ collective rotation:
$$ {\cal I} = \frac{8\pi}{3}\int^\infty_0\!\!r^2dr
\sin^2 F \left\{ f^2_\pi + \frac{1}{e^2} \left( F^{\prime 2}
+\frac{\sin^2 F}{r^2} \right)\right\},
\eqno(\mbox{\ref{mf}a})$$
and $c$ is the hyperfine splitting constant
which can be obtained by directly applying the techniques
developed in Ref.\cite{SMNR}:
\widetext
$$ \begin{array}{rl}
 c = &\hskip -3mm \displaystyle \left.\int^\infty_0\!\!
r^2dr \, \right\{ 2\omega_B^{} \left[ \textstyle
( |\varphi|^2 - \frac13 |\varphi^*_1|^2 - \frac13 |\varphi^*_2|^2 )
- \frac43 \cos^2\!(F/2) ( |\varphi|^2 - |\varphi^*_1|^2) \right] \\
& \hskip 3mm +\left.\frac13g_{_Q}\left[
 ( F' - \displaystyle \frac{\sin\!2F}{r}) |\varphi^*_2|^2
- \frac{1}{\sqrt2} \frac{\sin\!F}{r} ( 3\cos F + 1)
( \varphi_1^{*\dagger} \varphi^*_2
 + \varphi_2^{*\dagger} \phi_1 ) \right] \right\}. 
\end{array}\eqno(\mbox{\ref{mf}b})$$
\narrowtext
We note that we have also kept terms of the next to leading
order in $1/m_Q$. One can easily see that the hyperfine
constant $c$ is of order $1/m_Q$. The leading order terms
proportional to $\omega_B^{}$ vanish identically when the
radial functions of Eq. (\ref{IMlimit}a) are used.

According to the formula, the masses of heavy baryons containing
a single heavy quark have following hyperfine splittings:
\begin{eqnarray}\begin{array}{c}
\displaystyle M_{\Sigma_Q^*} - M_{\Sigma_Q}
= \frac{3}{2{\cal I}}\, c, \\
\displaystyle M_{\Sigma_Q} - M_{\Lambda_Q}
= \frac{1}{{\cal I}} \, ( 1 - c).
\end{array}\label{masssplit}
\end{eqnarray}
By eliminating $c$ from Eq. (\ref{masssplit}) we have
a model independent relation
\begin{eqnarray}
\frac13 ( 2 M_{\Sigma^*_Q} + M_{\Sigma_Q}) -  M_{\Lambda_Q}
= \frac23 ( M_\Delta - M_N ).
\label{Mrel}
\end{eqnarray}

With the experimental values $M^{\exp.}_{\Sigma_c}$($=2453$ MeV),
$M^{\exp.}_{\Lambda_c}$($=2285$ MeV) and
$M^{\exp.}_{\Lambda_b}$($=5641$ MeV), we predict the mass of
$\Sigma^*_c$ to be 2493 MeV and the averaged mass
$\overline{M}_{\Sigma_b}(\equiv\frac13(2M_{\Sigma^*_b}
+M_{\Sigma_b})$ 5836 MeV.  Since $c$ is of order $1/m_Q$,
the masses of $\Sigma_Q$ and of $\Sigma_Q^*$ are degenerate
in the infinite mass limit as the heavy quark symmetry implies
and Eq. (\ref{Mrel}) is reduced to
$M_{\Sigma_Q} - M_{\Lambda_Q} = \frac23(M_\Delta - M_N)$
as in Refs.\cite{JMW,MOPR,NRZ}.

Numerical results (Result I) on the heavy baryon masses are shown
in Table II. They are in rough agreement with the experimental
values. Result II is obtained by taking the two coupling constants
as free parameters. To fit the experimental masses of $\Lambda_c$ and
$\Sigma_c$, one should have $f_{_Q}/2M_{D^*}=-1.04$ and $g_{_Q}=-0.40$,
which implies that the heavy quark symmetric relation (\ref{rel})
is {\it strongly} broken in the charm sector. Note that as far as the
two coupling constants are related by Eq. (\ref{rel}), the hyperfine
constant is too small.

In order to improve the situation, one may consider higher order terms
in the $1/m_Q$ expansion or higher derivative terms of the pion fields.
As a guide line, we may use the Skyrme Lagrangian\cite{SMNR} with
the vector mesons included via the hidden gauge symmetry,
since in the strangeness sector
the heavy quark symmetry becomes no longer a good symmetry
but the $SU(3)$ chiral symmetry becomes rather a good symmetry.

Among many possible terms which will be discussed below,
the Wess-Zumino (WZ) term is known
to play the most important role in CK approach\cite{CK}:
\begin{eqnarray}
{\cal L}_{\mbox{\tiny WZ}}^{} =- \frac{iN_c}{4f_\Phi^2} B^\mu \left(
\Phi^\dagger D_\mu \Phi - (D_\mu \Phi)^\dagger \Phi \right).
\label{L1a}
\end{eqnarray}
Here, $N_c$ is the number of color, $f^{}_\Phi$ is the $\Phi$-meson
decay constant and $B_\mu$ is the topological baryon number current.
Although its role fades out in the heavy mass limit,
the WZ term should not be disregarded in the case of
finite quark masses.

In addition, there are other contributions of order $1/m_Q$ to
the binding potential. We introduce a typical $A \cdot A$ potential
in the Lagrangian as in CK approach, which is next to leading order
in the derivative of pion fields  and turns out to have
non-negligible effects in the strangeness sector:
\begin{eqnarray}
{\cal L}_{(2)} = - \Phi^\dagger A_\mu A^\mu \Phi.
\label{L1b}
\end{eqnarray}

Now we discuss the effects of the above terms in detail.
Let us write
\begin{eqnarray}
\delta {\cal L} = {\cal L}_{\mbox{\tiny WZ}}^{} + {\cal L}_{(2)}.
\end{eqnarray}
This additional Lagrangian modifies the equations of motion
for the pseudoscalar meson field $\Phi$ as
\begin{eqnarray}
(D_\mu D^\mu + M_\Phi^2)\Phi &=& f_{_Q} A^\mu \Phi_\mu^*
- \frac{2iN_c}{4f_\Phi^2}
B_\mu D^\mu \Phi \nonumber \\
&& - A_\mu A^\mu \Phi,
\end{eqnarray}
while those for the vector meson field $\Phi^*$ remain the same as
Eq. (\ref{14}). Consequently, the radial function of the
$k^P=\frac12^+$ eigenmodes is altered as
\widetext
\begin{eqnarray}
\begin{array}{l}
\displaystyle \varphi'' + \frac{2}{r} \varphi'
      + (\omega^2 - M_\Phi^2 -\frac{2}{r^2}) \varphi \\
\hskip 1cm = \frac12{f_{_Q}} ( a_1 + a_2 ) \phi_1
- \frac{1}{\sqrt2}f_{_Q} a_1 \phi_2 - ( 2\omega\lambda
- 2\upsilon(\upsilon - \displaystyle \frac{2}{r})
 - \textstyle \frac14 (3 a_1^2 + 2 a_1 a_2 + a_2^2 ) ) \varphi,
\end{array}
\end{eqnarray}
where
\narrowtext
\begin{eqnarray}
\lambda(r) = - \frac{N_c}{f_\Phi^2} \frac{1}{8\pi^2}
\frac{\sin^2 F}{r^2}F'.
\end{eqnarray}
The WZ term contributes to
the hyperfine splitting constant $c$ as
\begin{eqnarray}
\delta c = 2 \int^\infty_0\!\!\! r^2 dr |\varphi|^2 \lambda,
\end{eqnarray}
and to the normalization condition of Eq. (\ref{norm}) by the
same amount. Note that there is no direct contribution from
${\cal L}_{(2)}$ to this quantity.

We begin with the role of $\delta{\cal L}$ in the strangeness sector,
where the above model lagrangian does not work well as expected in
this sector. In Table III, we show the numerical results for strange
baryons obtained with the input parameters $f_\pi=64.5$ MeV,
$e=5.45$, $M_K=495$ MeV,
$M_{K^*}=892$ MeV and $f_{_Q}/2M_{K^*}=-\frac{1}{\sqrt2}=g_{_Q}$.
Note that the role of ${\cal L}_{(2)}$
in the binding energy is important ($\sim$ 80 MeV), while its effect
on the hyperfine constant $c$ is rather small.
The Wess-Zumino term plays a crucial role in the hyperfine constant $c$,
of which more than 80\% comes from the WZ term.
As shown in Fig. 2, the effect of the WZ term on the radial wavefunction
is also remarkable; with the WZ term, the vector meson contribution
to the bound states is  much suppressed compared with that of the
pseudoscalar meson.

In the charm sector the role of the Wess-Zumino term
in the heavy mass limit is weakened  as discussed in
Ref.\cite{MOPR}.  This results from the fact that the role of
vector mesons balances off that of pseudoscalar mesons in the
heavy mass limit. The decoupling of the WZ term in the heavy mass
limit is originally argued in Ref.\cite{ST,NRZ1}. In the chiral limit
where pseudoscalar mesons predominate, the WZ term is entirely
expressed in terms of these pseudoscalar mesons. To take into
account the WZ term in the finite mass region, we consider
the most characteristic expression of the WZ term\cite{CK,SMNR}
constructed in terms of pseudoscalar mesons and an adjustable
parameter $\gamma$. The parameter $\gamma$ contains the trace of
cancellation between contributions of the vector and of the
pseudoscalar mesons and should depend on $1/m_Q$.
That is, we take
\begin{eqnarray}
{\delta\cal L}'
 = \gamma {\cal L}_{\mbox{\tiny WZ}}^{} + \epsilon {\cal L}_{(2)}
\end{eqnarray}
with the same ${\cal L}_{\mbox{\tiny WZ}}$ and ${\cal L}_{(2)}$
as given by Eqs. (\ref{L1a},\ref{L1b}). The parameter $\epsilon$ has
the role of turning on and off the effect of ${\cal L}_{(2)}$.
Here, $f_\Phi$ is the $D$-meson decay constant $f_D^{}$,
which is known to be 1.8 times larger than the pion decay constant
$f_\pi$. Although the ${\cal L}_{(2)}$ plays a minor role for the
heavy flavors such as charm, we keep it to compare its effects in
the charm sector with those in the strangeness sector. In Fig. 3,
we present $\omega_B^{}$ and $c$ as a function of the mitigating
factor $\gamma$. The role of ${\cal L}_{(2)}$ is shown as narrow
stripes, with $\sim 30$~MeV effect on the energy and $\sim 0.04$
on the hyperfine constant. However, as we can see in Fig. 3,
the dependence of the mass spectrum on the parameter $\gamma$ is
not negligible. In order to fit the charmed baryon masses, we need
to have $\omega_B^{}=1416$ MeV and $c=0.16$. Then we have
\begin{eqnarray}
\begin{array}{l}
M_{\Lambda_c} = 2285 \mbox{ MeV}  \\
M_{\Sigma_c} = 2449 \mbox{ MeV} ,  \\
M_{\Sigma_c^*} = 2495 \mbox{ MeV}.
\end{array}
\end{eqnarray}
Also from Fig. 3, one can estimate the mitigating factors as
$\gamma\sim 0.25$ and $\epsilon = 1$, which reproduce the above mass
spectra and then we have $\sqrt{\langle r^2 \rangle_c} =
0.37$~fm. It implies that the increase in $f_\Phi$ alone is not
enough to take fully into account the role of the Wess-Zumino term
in the charm sector.

The dependence of $\gamma$ on meson masses can be derived by showing
how the Wess-Zumino term scales out as the mass increases.
We are not in a position to illustrate this dependence yet.
However, if we assume the dependence to be inversely proportional
to the $m_\Phi$, we find $\gamma\sim  0.5 \mbox{ GeV}/m_{\Phi}$.
Note the coincidence of the $\gamma$-factor ($\sim 0.25$)
with the meson mass ratio, $m_K/m_D$.

\section{Summary and Conclusion}
In this work, we have investigated the mass spectrum of heavy
baryons containing a single heavy quark in the bound state
approach of Skyrme model. To this end, we have worked
with the heavy meson Lagrangian of Ref.\cite{Yan} which
includes the  $1/m_Q$ order terms. The large binding energy
obtained in the infinite mass limit is lowered by introducing
$1/m_Q$ corrections. The binding energy is changed from
$\sim 800$ MeV to $\sim 500$ MeV for $D(D^*)$ mesons and to
$\sim 600$ MeV for the $B(B^*)$. The effect may be crucial for
the loosely bound exotic states such as ``pentaquark"
baryons\cite{Penta,OP}. However, due to the realization of the
heavy quark symmetry, the hyperfine splitting constant comes out
too small compared with the experimental one. For example, we
get $c=0.05$ for charmed baryons while it should be $\sim 0.14$
to reproduce the experimental masses. To resolve this problem, we
introduce the WZ term in a mitigated form, which is known to have
a crucial role in the strangeness  sector. To reproduce the
experimental masses for charmed baryons, its strength should
be weakened by a factor of $4$.

\acknowledgments
We would like to thank Mannque Rho for having brought us this
problem and for enlightening discussions. We also would like
to thank H. Jung for reading the manuscript and useful comments.
Also, Y. Oh is grateful to N. N. Scoccola for useful discussions.
This work is supported in part by the Korea Science and Engineering
Foundation through the SRC program and by CNU Research and
Scholarship Fund.

\vskip 5cm

\begin{figure}
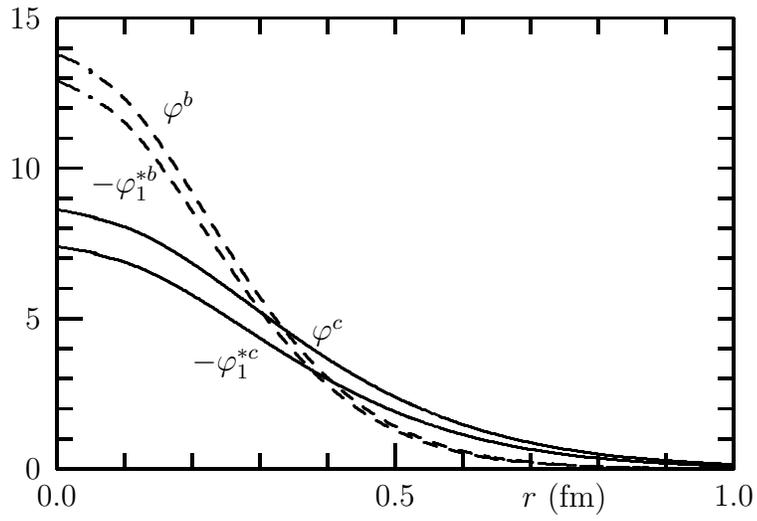

\caption{$\varphi(r)$ and $\varphi_1^*(r)$ for $Q=c$ (solid) and
 $b$ (dashed). $\varphi_2^*(r)$ is nearly equal to $\protect\sqrt{2}
 \varphi_1^*(r)$ for both cases.}
\end{figure}

\begin{figure}
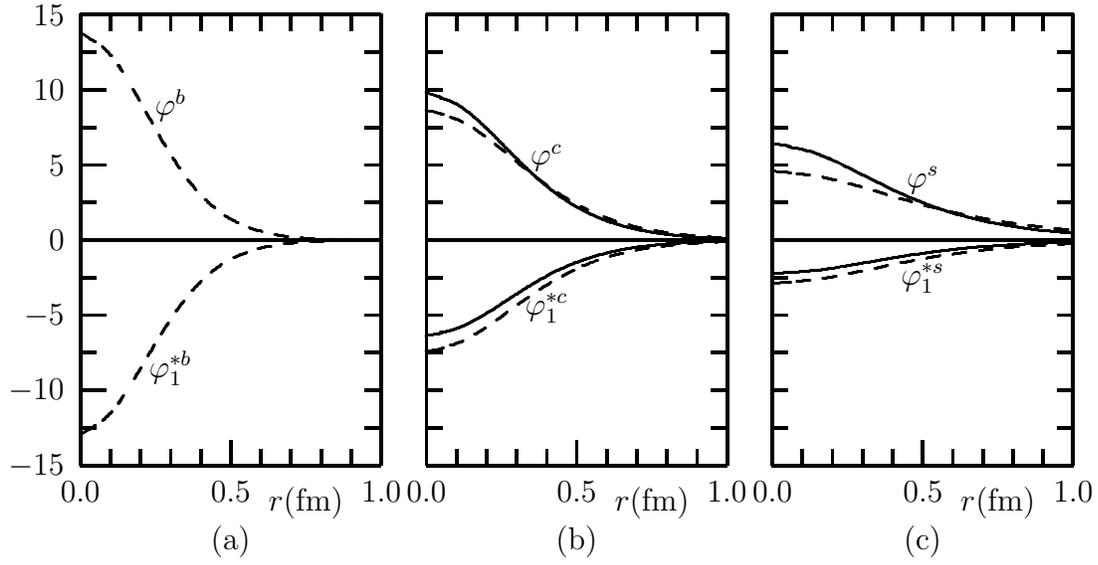

\caption{$\varphi(r)$ and $\varphi^*_1(r)$ for (a) $B$ and $B^*$,
(b) $D$ and $D^*$, (c) $K$ and $K^*$ with (solid) and without
(dashed) the Wess-Zumino term.
Each fields are normalized as $\int dr r^2 | \varphi |^2 = 1$.}
\end{figure}

\begin{figure}
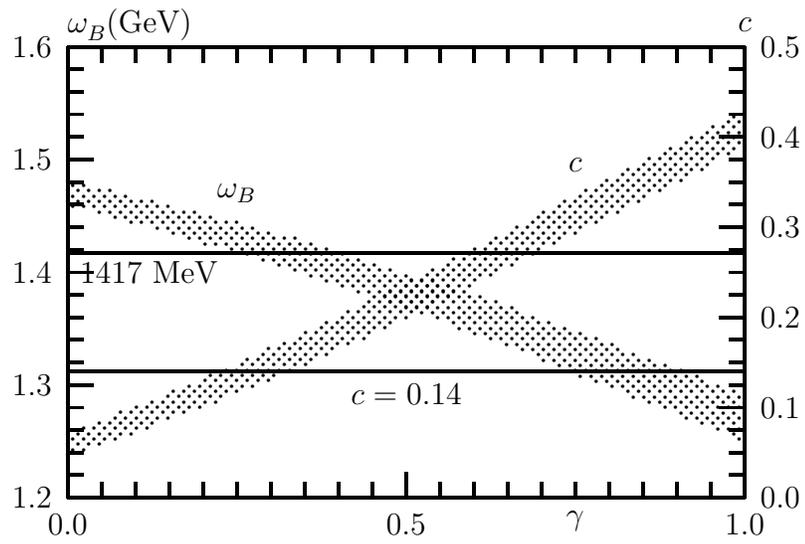

\caption{$\omega_B^{}$ and $c$ vs. $\gamma$ obtained for
         charmed baryons.}
\end{figure}

\newpage

\widetext

\begin{table}
\caption{Summary on the input parameters and the numerical results
         on the bound state }
\begin{tabular}{cccrrrcrccc} \hline
$Q$  & $f_\pi^{a)}$ & $e^{c)}$ & \multicolumn{1}{c}{$M_\Phi^{a)}$}
& $M_{\Phi^*}^{a)}$ & \multicolumn{1}{c}{\ $f_{_Q}^{a)}$}
& \ $g_{_Q}^{c)}$
& \multicolumn{1}{c}{$\;\omega^{a)}_{B}$}
& $\sqrt{\langle r^2\rangle}^{b)}$ & $c^{c)}$
& $\varphi^*_1(0)/\varphi(0)$ \\
\hline
$c$ & 64.5 & 5.45 & 1872 & 2010 & $-$3016
 & $-$0.75 & 1481 & 0.39 & 0.05 & $-$0.828 \\
$b$ & 64.5 & 5.45 & 5275 & 5325 & $-$7988
 & $-$0.75 & 4722 & 0.29 & 0.02 & $-$0.932 \\
\hline
 & \multicolumn{10}{c}{$^{a)}$ in MeV unit, $^{b)}$ in fm unit, and
$^{c)}$ dimensionless quantities. }
\end{tabular}
\end{table}

\begin{table}
\caption{Numerical results on the heavy baryon masses.}
\begin{tabular}{ccccccccc}
\hline
$Q$ &   & $f_{_Q}/2M_{\Phi^*}$ & $g_{_Q}$
& $\omega_B^{a)}$ & $c$
& $M_{\Lambda_Q}^{a)}$ & $M_{\Sigma_Q}^{a)}$
& $M_{\Sigma^*_Q}^{a)}$ \\
\hline
 & exp.$^{b)}$ &   &   &  &   & 2285 & 2453 & ---  \\
$c$ &  I   & $-$0.75 & $-$0.75
& 1481 & 0.05 & 2348 & 2535 & 2548 \\
    & II   & $-$1.04 & $-$0.40
& 1419 & 0.14 & 2287 & 2454 & 2497 \\
\hline
$b$ & exp.$^{b)}$ &     &      &      &
& 5641 &  --- & ---  \\
    &  I   & $-$0.75 & $-$0.75 & 4722
& 0.02 & 5589 & 5781 & 5786 \\
\hline
 & \multicolumn{8}{c}{$^{a)}$ in MeV unit,
$^{b)}$ Particle Data Group\cite{PDG}.}
\end{tabular}
\end{table}

\begin{table}
\caption{WZ term and strange baryon masses.}
\begin{tabular}{cccccccc}
\hline
${\cal L}_{\mbox{\tiny WZ}}$ & ${\cal L}_{(2)}$ & $\omega^{a)}_B$ & $c$
 & $M^{a)}_{\Lambda}$ & $M^{a)}_{\Sigma}$ & $M^{a)}_{\Sigma^*}$
 & $\sqrt{\langle r^2\rangle}^{b)}$ \\
\hline
off & off & 389 & 0.098 & 1257 & 1433 & 1462 & 0.62 \\
off & on  & 291 & 0.148 & 1160 & 1326 & 1369 & 0.56 \\
on  & off & 191 & 0.717 & 1095 & 1151 & 1361 & 0.41 \\
on  & on  & 109 & 0.791 & 1022 & 1063 & 1295 & 0.39 \\
\multicolumn{2}{c}{exp.} & & & 1116 & 1192 & 1385 & -- \\
\hline
\multicolumn{8}{c}{ $^{a)}$ in MeV unit, $^{b)}$ in fm unit.}
\end{tabular}
\end{table}
\clearpage

\
\vspace{7cm}
\begin{figure}
\beginpicture
\setcoordinatesystem units <1cm,1cm> point at 3 2
\setplotarea x from 0 to 15, y from 0 to 8
\put {{\bf Fig.1 :} $\varphi(r)$ and $\varphi_1^*(r)$ for
   $Q$=$c$ (solid) and $b$ (dashed).}  at 7.5 0.5
\put{$\varphi_2^*(r)$ is nearly equal to $\sqrt2\varphi_1^*(r)$ for both
cases.} at 7.75 0
\setcoordinatesystem units <9cm,0.40cm> point at 0 0
\setplotarea x from 0 to 1.0, y from 0 to 15.0
\linethickness=1pt
\thicklines
\axis bottom ticks in
   numbered from 0 to 1.0 by 0.5
   unlabeled short from 0 to 1.00 by 0.1 /
\axis left ticks in
   numbered  from 0 to 15 by 5
   unlabeled short from 1 to 14 by 1 /
\axis top ticks in
   unlabeled short from 0.1 to 0.9 by 0.1 /
\axis right ticks in
   unlabeled short from 1 to 14 by 1 /
\put {$r$ (fm)} at 0.75 -1
\setquadratic
\inboundscheckon
\setplotsymbol ({.})
\plot
    .00120 8.6307   .04918 8.4075    .05517 8.3504   .07317 8.2363
    .09116 8.1247   .10916 7.9766    .12715 7.7979   .14514 7.5931
    .16314 7.3658   .18113 7.1195    .19912 6.8574   .21711 6.5827
    .23511 6.2982   .27109 5.7115    .30708 5.1181   .34306 4.5357
    .37905 3.9783   .41503 3.4561    .45102 2.9760   .48701 2.5418
    .52299 2.1548   .52899 2.0949    .56497 1.7619   .60096 1.4727
    .63695 1.2241   .67293 1.0122    .70892 .83314   .74490 .68289
    .78089 .55763   .87685 .31985    1.0088 .14495 /
\put {$\varphi^c_{}$} at 0.4 4.5
\plot
    .00120 7.3999   .04918 7.2171   .05517 7.1595   .07317 7.0490
    .09116 6.9466   .10916 6.8115   .12715 6.6494   .14514 6.4641
    .16314 6.2590   .18113 6.0373   .19912 5.8021   .21711 5.5560
    .23511 5.3020   .27109 4.7805   .30708 4.2565   .34306 3.7456
    .37905 3.2603   .41503 2.8092   .45102 2.3979   .48701 2.0291
    .52299 1.7033   .52899 1.6531   .56497 1.3759   .60096 1.1374
    .63695 .93444   .67293 .76331   .70892 .62024   .74490 .50155
    .78089 .40378   .87685 .22224   1.0088 .09438 /
\put {$-\varphi_1^{*c}$} at 0.25 3.5
\setdashes
\plot
    .00120 13.786   .04318 13.363   .04918 13.186
    .06717 12.949   .08516 12.646   .10316 12.253    .12115 11.788
    .13914 11.263   .15714 10.692   .17513 10.086    .19312 9.4557
    .24110 7.7340   .27709 6.4795   .31308 5.3152    .34906 4.2743
    .38505 3.3738   .42103 2.6170   .45702 1.9973    .49300 1.5014
    .52899 1.1129   .56497 .81418   .60096 .58849    .63695 .42062
    .67293 .29753   .76889 .11301   .83487 .056235   .90084 .02738
    .96681 .01309  1.0328 .00617 /
\put {$\varphi^b_{}$} at 0.18 12
\plot
    .00120 12.923   .04318 12.465   .04918 12.359    .06717 12.131
    .08517 11.842   .10316 11.469   .12115 11.027    .13914 10.530
    .15714 9.9887   .17513 9.4148   .19312 8.8187    .24110 7.1935
    .27709 6.0125   .31308 4.9192   .34906 3.9446    .38505 3.1040
    .42103 2.3998   .45702 1.8250   .49300 1.3668    .52899 1.0091
    .56497 .73519   .60096 .52905   .63695 .37636    .67293 .26490
    .76889 .09914   .83487 .04877   .90084 .02344    .96681 .01105
    1.0328 .00513 /
\put {$-\varphi^{*b}_1$} at 0.10 9.5
\endpicture
\end{figure}
\clearpage
\ \vskip 5cm
\begin{figure} 
\beginpicture
\setcoordinatesystem units <1cm,1cm> point at 1 2.5
\setplotarea x from 0 to 15, y from 0 to 8.5
\put {{\bf Fig.2 :} $\varphi(r)$ and $\varphi^*_1(r)$ for (a) $B$ and $B^*$,
(b) $D$ and $D^*$, (c) $K$ and $K^*$ } at 7.5 0.5
\put{ with (solid)
and without (dashed) the Wess-Zumino term.} at 7.75 0.
\put{Each fields are normalized as
$\int dr r^2 | \varphi |^2 = 1$.} at 7.75 -0.5
\setcoordinatesystem units <4cm,0.2cm> point at 0 -15
\linethickness=1pt
\thicklines
\setplotarea x from 0 to 1.0, y from -15 to 15
   \axis bottom ticks in
     numbered from 0 to 1.0 by 0.5
     unlabeled short quantity 11 /
\axis left ticks in
   numbered from -15 to 15 by 5 unlabeled short quantity 13 /
\axis top ticks in unlabeled short quantity 11 /
\axis right ticks in unlabeled short quantity 13  /
\setsolid
\setplotsymbol ({.})
\setquadratic
\putrule from 0 0 to 1 0
\put {$r$(fm)} at 0.75 -17.5
\put {(a)} at 0.5 -20
\inboundscheckon
\setdashes\plot
    .00120  13.801    .04318  13.276    .05518  13.102
    .06717  12.962    .07917  12.770    .09116  12.535
    .10316  12.264    .11515  11.960    .12715  11.628
    .13914  11.272    .17513  10.092    .21112  8.8158
    .24710  7.5225    .28309  6.2782    .31907  5.1312
    .35506  4.1120    .39104  3.2350    .42703  2.5015
    .46302  1.9035    .49900  1.4269    .56497  .81201
    .63095  .44361    .69692  .23392    .76290  .11963
    .82887  .05960    .89484  .02904  
    1.0268  .00654 /
\put {$\varphi^b$} at 0.3 9
\plot
    .00120    -12.905 .043183     -12.418 .055178     -12.247
    .06717     -12.113 .079169     -11.931 .091164     -11.709
    .10316      -11.452 .11515      -11.164 .12715      -10.850
    .13914      -10.513 .17513      -9.3987 .21112      -8.1949
    .24710      -6.9780 .28309      -5.8100 .31907      -4.7361
    .35506      -3.7846 .39104      -2.9682 .42703      -2.2877
    .46302      -1.7347 .49900      -1.2955 .56497      -.73180
    .63095      -.39652 .69692      -.20718 .76290      -.10489
    .82887      -.05166 .89484      -.02486 
    1.0268      -.00544 /
\put {$\varphi^{*b}_1$} at 0.3 -8.5
\setsolid
\setcoordinatesystem units <4cm,0.2cm> point at -1.15 -15
\setplotarea x from 0 to 1.0, y from -15 to 15
   \axis bottom ticks in
     numbered from 0 to 1.0 by 0.5
     unlabeled short quantity 11 /
\axis left ticks in
   unlabeled short quantity 13 /
\axis top ticks in unlabeled short quantity 11 /
\axis right ticks in unlabeled short quantity 13 /
\putrule from 0 0 to 1 0
\put {$r$(fm)} at 0.75 -17.5
\inboundscheckon
\put {(b)} at 0.5 -20
\setsolid \plot
    .00120       9.8172 .04318       9.5371 .04918       9.4243
    .05517       9.4095 .06117       9.3851 .06717       9.3524
    .07317       9.3124 .08516       9.2136 .09716       9.0927
    .10916       8.9527 .12115       8.7955 .13315       8.6231
    .14514       8.4372 .15714       8.2392 .16913       8.0307
    .20512       7.3559 .24110       6.6351 .27709       5.9004
    .31308       5.1787 .34906       4.4908 .38505       3.8516
    .42103       3.2704 .45702       2.7517 .49300       2.2963
    .52899       1.9020 .59496       1.3238 .66094       .90427
    .72691       .60827 .79288       .40407 .85886       .26575
    .92483       .17340 .99080       .11246 1.0568       .07260 /
\put {$\varphi^c$} at 0.4 5.5
\plot
    .00120      -6.3616 .04318      -6.2120 .04918      -6.1111
    .05517      -6.1027 .06117      -6.0880 .06717      -6.0681
    .07317      -6.0436 .08516      -5.9826 .09716      -5.9077
    .10916      -5.8205 .12115      -5.7225 .13315      -5.6148
    .14514      -5.4983 .15714      -5.3740 .16913      -5.2429
    .20512      -4.8166 .24110      -4.3583 .27709      -3.8879
    .31308      -3.4225 .34906      -2.9758 .38505      -2.5579
    .42103      -2.1756 .45702      -1.8324 .49300      -1.5294
    .52899      -1.2661 .59496      -.87815 .66094      -.59579
    .72691      -.39665 .79288      -.25983 .85886      -.16789
    .92483      -.10725 .99080      -.06787 1.0568      -.04262 /
\setdashes
\plot
    .00120 8.6307   .04918 8.4075    .05517 8.3504   .07317 8.2363
    .09116 8.1247   .10916 7.9766    .12715 7.7979   .14514 7.5931
    .16314 7.3658   .18113 7.1195    .19912 6.8574   .21711 6.5827
    .23511 6.2982   .27109 5.7115    .30708 5.1181   .34306 4.5357
    .37905 3.9783   .41503 3.4561    .45102 2.9760   .48701 2.5418
    .52299 2.1548   .52899 2.0949    .56497 1.7619   .60096 1.4727
    .63695 1.2241   .67293 1.0122    .70892 .83314   .74490 .68289
    .78089 .55763   .87685 .31985    1.0088 .14495 /
\plot
    .00120 -7.3999   .04918 -7.2171   .05517 -7.1595   .07317 -7.0490
    .09116 -6.9466   .10916 -6.8115   .12715 -6.6494   .14514 -6.4641
    .16314 -6.2590   .18113 -6.0373   .19912 -5.8021   .21711 -5.5560
    .23511 -5.3020   .27109 -4.7805   .30708 -4.2565   .34306 -3.7456
    .37905 -3.2603   .41503 -2.8092   .45102 -2.3979   .48701 -2.0291
    .52299 -1.7033   .52899 -1.6531   .56497 -1.3759   .60096 -1.1374
    .63695 -.93444   .67293 -.76331   .70892 -.62024   .74490 -.50155
    .78089 -.40378   .87685 -.22224   1.0088 -.09438 /
\put {$\varphi_1^{*c}$} at 0.4 -4.5
\setsolid
\setcoordinatesystem units <4cm,0.2cm> point at -2.3 -15
\setplotarea x from 0 to 1.0, y from -15 to 15
   \axis bottom ticks in
     numbered from 0 to 1.0 by 0.5
     unlabeled short quantity 11 /
\axis left ticks in
   unlabeled short quantity 13 /
\axis top ticks in unlabeled short quantity 11 /
\axis right ticks in unlabeled short  quantity 13 /
\putrule from 0 0 to 1 0
\put {$r$(fm)} at 0.75 -17.5
\inboundscheckon
   \put {(c)} at 0.5 -20
\setsolid
\plot
    .00120       6.4005 .04218       6.3068 .06117       6.1711
    .06717       6.1606 .07317       6.1463 .07916       6.1287
    .09116       6.0846 .10316       6.0303 .11515       5.9671
    .12715       5.8959 .13914       5.8175 .15114       5.7325
    .16314       5.6417 .17513       5.5455 .18713       5.4445
    .19912       5.3394 .23511       5.0035 .27109       4.6473
    .30708       4.2822 .34306       3.9180 .37905       3.5626
    .41503       3.2220 .45102       2.9004 .48701       2.6005
    .52299       2.3238 .55898       2.0707 .62495       1.6666
    .69092       1.3343 .75690       1.0648 .82287       .84835
    .88884       .67556 .95482       .53821 1.0208       .42926 /
\put {$\varphi^s$} at 0.5 4
\plot
    .00120      -2.2141 .04218      -2.1706 .06117      -2.1369
    .06717      -2.1337 .07317      -2.1292 .07916      -2.1235
    .09116      -2.1093 .10316      -2.0917 .11515      -2.0711
    .12715      -2.0477 .13914      -2.0219 .15114      -1.9939
    .16314      -1.9639 .17513      -1.9320 .18713      -1.8984
    .19912      -1.8633 .23511      -1.7505 .27109      -1.6298
    .30708      -1.5049 .34306      -1.3790 .37905      -1.2550
    .41503      -1.1349 .45102      -1.0204 .48701      -.91276
    .52299      -.81261 .55898      -.72034 .62495      -.57180
    .69092      -.44880 .75690      -.34888 .82287      -.26897
    .88884      -.20592 .95482      -.15673 1.0208      -.11874 /
\put {$\varphi^{*s}_1$} at 0.5 -2.5
\setdashes
\plot
    .00120       4.5936 .05517       4.5026 .06117       4.4527
    .08516       4.4296 .11515       4.3648 .15114       4.2501
    .18713       4.1042 .22311       3.9340 .25910       3.7455
    .29508       3.5441 .33107       3.3349 .36705       3.1223
    .40304       2.9101 .46901       2.5322 .53499       2.1801
    .60096       1.8619 .66693       1.5808 .73291       1.3367
    .79888       1.1274 .86485       .94971 .93083       .79984
    .99680       .67399 1.0628       .56859 /
\plot
    .00120      -2.8501 .05517      -2.8066 .06117      -2.7553
    .08516      -2.7339 .11515      -2.6820 .15114      -2.5931
    .18713      -2.4821 .22311      -2.3542 .25910      -2.2142
    .29508      -2.0666 .33107      -1.9153 .36705      -1.7636
    .40304      -1.6144 .46901      -1.3546 .53499      -1.1197
    .60096      -.91403 .66693      -.73841 .73291      -.59130
    .79888      -.47002 .86485      -.37131 .93083      -.29185
    .99680      -.22847 1.0628      -.17830 /
\endpicture
\end{figure}
\clearpage

\ \vskip 5cm
\begin{figure} 
\beginpicture
\setcoordinatesystem units <1cm,1cm> point at 3 2
\setplotarea x from 0 to 15, y from 0 to 8
\put {{\bf Fig.3 :} $\omega_B^{}$ and $c$ vs. $\gamma$ obtained
for charmed baryons.} at 7.5 0.5
\setcoordinatesystem units <9.0cm,12.0cm> point at -0. 0.
\linethickness=1pt \thicklines
\setplotarea x from 0 to 1.0, y from 0 to 0.5
\axis bottom ticks in
   numbered from 0 to 1.0 by 0.5
   unlabeled short quantity 21 /
\axis right ticks in
   numbered from 0 to 0.5 by 0.1
   unlabeled short quantity 21 /
\put {$\gamma$} at 0.75 -0.025
\put {$c$} at 1.0 0.525
\put {$\omega_B^{}$(GeV)} [l] at 0 0.525
\putrule from 0 0.14 to 1 0.14
\put {$c$} at 0.75 0.37
\put {$c=0.14$} at 0.5 0.115
\setshadesymbol <z,z,z,z> ({\bf$\cdot$})
\setshadegrid span <2pt>
\setquadratic \vshade
0.000    0.046   0.072  0.125    0.078   0.108 0.250    0.114   0.148
0.375    0.154   0.192 0.500    0.198   0.238 0.625    0.244   0.286
0.750    0.291   0.335 0.875    0.339   0.383 1.000    0.387   0.429 /
\setcoordinatesystem units <9.0cm,15cm> point at -0. 1.2
\setplotarea x from 0 to 1.0, y from 1.2 to 1.6
\axis top ticks in
   unlabeled short from 0.05 to 0.95 by 0.05 /
\axis left ticks in
   numbered from 1.2 to 1.6 by 0.1
   unlabeled short quantity 21 /
\putrule from 0 1.417 to 1 1.417
\put {$\omega_B$} at 0.25 1.470
\put {1417 MeV} at 0.12 1.4
\vshade
0.000   1.458   1.481 0.125   1.438   1.463 0.250   1.416   1.444
0.375   1.392   1.423 0.500   1.366   1.399 0.625   1.339   1.375
0.750   1.309   1.348 0.875   1.279   1.320 1.000   1.248   1.291 /
\endpicture
\end{figure} 


\begin{references}
\bibitem{fnt}
we combine first the spin basis and the orbital angular momentum
basis to the total spin ($\vec{J}=\vec{S}+\vec{L}$) basis
and then combine the isospin isospin.
Then, $\vec{\cal Y}_{\frac12,+,\pm\frac12}^{(1)}(\hat{r})$ and
$\vec{\cal Y}_{\frac12,+,\pm\frac12}^{(2)}(\hat{r})$ correspond
to $J=0$ and $J=1$ states, respectively. One may obtain the vector
spherical harmonics in the other ways; for example, by combining
first the isospin basis and the orbital angular momentum basis
to $\vec{\Lambda}(=\vec{I}+\vec{L})$ basis and then the spin basis,
which leads us to
\begin{eqnarray*}
\vec{\cal Y}_{\frac12,+,\pm\frac12}^{\prime(1)}(\hat{r})
&=& \frac{1}{\sqrt{12\pi}} \vec\tau \cdot \hat r \vec{\tau}
\chi_{\pm} \nonumber \\
&=&
\sqrt{\frac13}\vec{\cal Y}_{\frac12,+,\pm\frac12}^{(1)}
+\sqrt{\frac23}\vec{\cal Y}_{\frac12,+,\pm\frac12}^{(2)},
  \nonumber \\
\vec{\cal Y}_{\frac12,+,\pm\frac12}^{\prime(2)}(\hat{r})
&=& \frac{1}{\sqrt{24\pi}} (\vec\tau \cdot \hat r\hat{r}-3\hat{r})
 \chi_{\pm} \nonumber \\
&=& -\sqrt{\frac23}\vec{\cal Y}_{\frac12,+,\pm\frac12}^{(1)}
+\sqrt{\frac13}\vec{\cal Y}_{\frac12,+,\pm\frac12}^{(2)}.
\end{eqnarray*}
They are $\Lambda=\frac12$ and $\Lambda=\frac32$ states, respectively.
\end{references}
\end{document}